\documentclass{eas}
\usepackage{graphicx}
\usepackage{natbib}
\begin{document}

%%-----------------------------
%%      the top matter
%%-----------------------------
\title{Triggered Star Formation}
\author{Bruce G. Elmegreen}\address{IBM T. J. Watson Research Center, 1101 Kitchawan Road, Yorktown
Heights, New York 10598 USA, bge@us.ibm.com}
\begin{abstract}
Triggered star formation in bright rims and shells is reviewed. Shells
are commonly observed in the Milky Way and other galaxies, but most
diffuse shells seen in HI or the infrared do not have obvious triggered
star formation. Dense molecular shells and pillars around HII regions
often do have such triggering, although sometimes it is difficult to
see what is triggered and what stars formed in the gas before the
pressure disturbances. Pillar regions without clear age gradients could
have their stars scattered by the gravity of the heads. Criteria and
timescales for triggering are reviewed.  The insensitivity of the
average star formation rate in a galaxy to anything but the molecular
mass suggests that triggering is one of many processes that lead to
gravitational collapse and star formation.
\end{abstract}
\maketitle

\section{Introduction: Large-Scale Shells}

\begin{figure}[b]
% \vspace*{-2.0 cm}
\begin{center}
 \includegraphics[width=3.in]{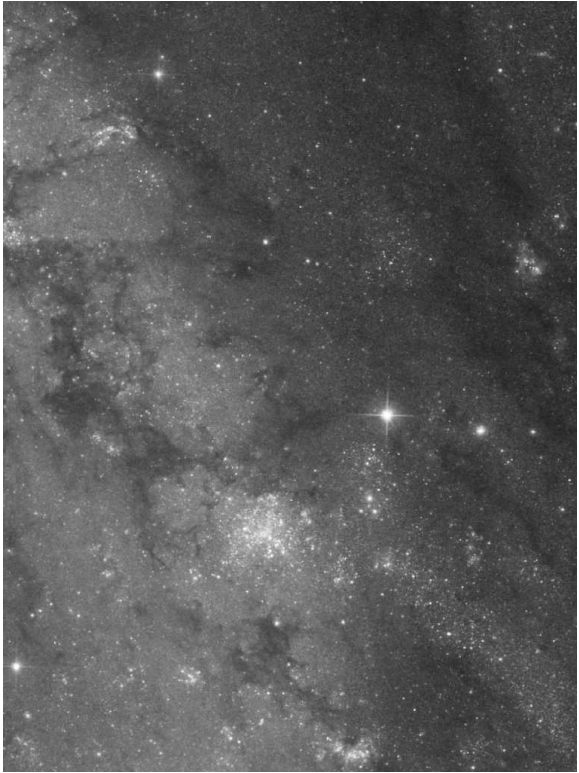}
% \vspace*{-1.0 cm}
\caption{Region of the galaxy M51 viewed with the ACS camera of HST,
showing bubbles.}\label{m51}
\end{center}
\end{figure}

High resolution images of nearby spiral galaxies show large dust and
gas bubbles in the spiral arms (Fig. \ref{m51}; see also Fig. 3 in
Lecture 2). Their radii are often larger than the gas scale heights, so
these are unlikely to be spheres; they are more like rings in the disk.
There are also feathers, comets, and other fine-scale dust structures
in optical images -- all indicating recent dynamical processes.
Sometimes there are small bubbles inside large bubbles, and there are
generally more bubbles near the spiral arms than in the interarm
regions.  The interarm also contains dust streamers, and many of these
look like old bubbles left over from more active times in the arms.

Gould's Belt contains the Local Bubble, studied recently by
\cite{lallement03} using Na I absorption clouds near the Sun. This
local bubble is also a source of diffuse x-ray emission from hot gas
\citep{snowden98}. Presumably the energy came from hot stars in the
Sco-Cen association \citep{breit06}.

Large bubbles in the Milky Way are also evident from IR maps of the
sky. \cite{kon07} used IRAS $60\mu$m and $100\mu$m images to identify
Milky Way bubbles. They reported that the bubble volume filling factor
in the inner galaxy is around 30\%, and in the outer galaxy it is
around 5\%. \cite{ehl05} catalogued HI shells in the Milky Way using
the Leiden-Dwingeloo Survey. They found a volume filling factor of 5\%,
a mean age of 8.4 Myr, and a ratio of age to filling factor equal to
170 Myr.  This latter timescale is the time for the whole interstellar
volume (not mass) to be cycled through one or another HI shell. This is
10 times faster than the time for molecular gas to be converted into
stars. Since the shells observed by Ehlerov\'a \& Palou\v s are atomic,
this ISM processing would seem to be independent of the molecular cloud
population.  The GMCs have a low volume filling factor and the HI
shells occupy the space between them. A significant fraction of the HI
shell mass can come from GMC disruption in the inner galaxy where the
molecular fraction is high.

The LMC is also filled with large shells \citep{goudis78}. The largest,
LMC4, has no obvious cluster or OB association in the center, although
there are A-type stars suggesting a $\sim30$ Myr old age \citep{efe98}.
It has pillars at the edge with young star formation, and an arc of
young stars without gas in the center, called Constellation III
\citep{mckibben53}. \cite{yamaguchi01} studied the star formation in
this region, pointing out GMCs and young clusters all along the edge of
the shell and suggesting these were triggered.

IC 10 is another small galaxy filled with HI holes and shells.
\cite{wilcots98} found H$\alpha$ at the edges of the shells and
discussed these young regions as triggered star formation. Similarly,
the small galaxy IC 2574 has a giant shell with an old central cluster
and triggered young stars on edge \citep{walter99, connon05}.

\section{Shell Expansion}

Expanding shells are most commonly made by stellar pressures in the
form of HII regions, supernovae, and winds. If we write the expansion
speed as $dR/dt\sim(P/\rho)^{1/2}$ for an isothermal shock, then the
radius varies as a power law in time if the pressure $P$ is a function
of radius $R$ and the density $\rho$ is uniform. For an HII region,
$P=2.1nkT$ where $n= (3S/4\pi R^3 \alpha)^{1/2}$ for ionizing
luminosity $S$ in photons per second and recombination rate $\alpha$ to
all but the ground state. Then $P\propto R^{-3/2}$. For supernovae,
$P\sim3E/4\pi R^3$ for the energy conserving, non-radiative, phase. For
a wind, $P\sim3E(t)/4\pi R^3$, where the energy increases with time as
$E=Lt$.

These three pressure-radius relations give three different radius-time
expansion laws, $dR/dt\propto R^{-3/4}$ gives $R\propto t^{4/7}$ for a
Str\"omgren sphere, $dR/dt\propto R^{-3/2}$ gives $R\sim t^{2/5}$ for
the Sedov phase of a supernova, and $dR/dt\propto t^{1/2}R^{-3/2}$
gives $R\propto t^{3/5}$ for a steady wind or continuous energy supply
from multiple supernovae in an OB association \citep{castor75}.

There are many complications to these solutions. External pressure is
always present, slowing down the bubbles. External pressure $P_{\rm
ext}$ enters the expression as $dR/dt = ([P-P_{\rm ext}]/\rho)^{1/2}$
with $P_{\rm ext}\sim$constant. The solution is not a power law in this
case. A second complication is the momentum in the moving shell. When
this is important, the equation of expansion is really $d(4\pi R^3 v
\rho/3)/dt = 4\pi R^2 (P-P_{\rm ext})$. Shell momentum makes the shell
move faster at a given radius than in the case without momentum. There
are also diverse shock jump conditions depending on the importance of
magnetic fields and the equation of state for the shocked gas, such as
adiabatic or isothermal, or whether the full energy equation is used to
determine the post-shock temperature.

We can see how important external pressure is to these solutions by
finding the fraction of shells that are at a pressure significantly
above the external value. As noted above, each source has solutions
$R(t)$ and $P(R)$, which can be re-written into a solution for pressure
versus time, $P(t)$. Thus there is a relation for the volume as a
function of pressure, $V(P)$. For a constant rate $n_0$ of making
bubbles, $n(P)dP=n_0dt$. Therefore $n(P)\propto dt/dP$. The volume
filling factor is $f(P)=n(P)V(P)$. Now we see that for HII regions,
$f(P)\propto P^{-4.17}$; for winds, $f(P)\propto P^{-4.5}$, and for
supernovae, $f(P)\propto P^{-5.2}$. For all of these, approximately,
$f(P)dP\propto AP^{-4.5}dP$ for some constant $A$. If all of the volume
is filled, then $1=\int f(P)dP$, and the average pressure is related to
the minimum pressure as $P_{\rm ave}=1.4P_{\rm min}$, which means
$f(P)=1.15(P/P_{\rm ave})^{-4.5}/P_{\rm ave}$. Thus, the probability
that any of these regions has a pressure exceeding 10 times the
average, $f(P>10P_{\rm ave})$, is $0.31\times 0.1^{3.5}\sim10^{-4}$;
similarly, $f(P>2P_{\rm ave})\sim0.03$. Evidently, most pressure bursts
from HII regions, winds and supernovae are within twice the average ISM
pressure for most of their lives. Therefore the external pressure is
important for them. \cite{kim01} ran numerical simulations of the ISM
and found that most of the time, the pressure stayed within a factor of
2 of the average value.

The probability distribution function for pressure also suggests that
the largest pressure bursts are close-range and short-lived. Thus
significant over-pressures from stellar sources are most likely to
occur close to those stars, as in an adjacent cloud. Most of the giant
IR and HI shells discussed above are drifting by momentum conservation.

\begin{figure}[b]
% \vspace*{-2.0 cm}
\begin{center}
 \includegraphics[width=3.in]{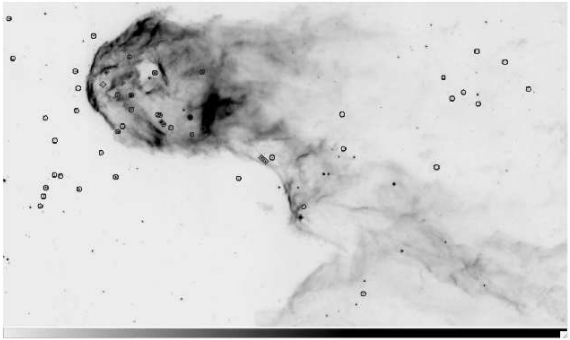}
% \vspace*{-1.0 cm}
\caption{A pillar in IC 1396 viewed at 8$\mu$m with the Spitzer Space
Telescope. Class I sources (the youngest) are identified by diamond
shapes. There are three at the front of the head, one near the back
part of the head, two on the lower part of the pillar and another in a
shelf nearby (from \cite{reach09}.}\label{reach}
\end{center}
\end{figure}

\section{Triggering: Bright Rimmed Clouds and Pillars}

HII regions interact with their neighboring molecular clouds by pushing
away the lower density material faster than the higher density cloud
cores. This leads to bright rims and pillars. Most HII regions contain
these shapes, as they are commonly observed in Hubble Space Telescope
images of nebulae. Triggered star formation in the dense heads of
pillars has been predicted \cite[e.g.,][]{klein80} and observed for
many years \citep[e.g.,][]{sugitani89}. Here we review some recent
observations.

Recent simulations of bright rim and pillar formation are in
\cite{mellema06}, \cite{miao06}, and \cite{grit09}. In a large HII
region, there can be many bright rims with star formation in them. A
good example is 30 Dor in the LMC, which has bright-rims that look like
they have triggered star formation in many places \citep{walborn02}.

IC 1396 is an HII region with a shell-like shape. The radius is 12 pc,
and the expansion speed is 5 km s$^{-1}$, making the expansion time 2.5
Myr \citep{patel95}.  The shell contains several bright rims and
pillars around the edge that all point to the sources of radiation. In
optical light, few embedded stars can be seen, but in the infrared
there are often embedded stars.

The stellar content of the large pillar in IC 1396 has been studied by
Reach et al. (2009; see Figure \ref{reach}). Several Class I protostars
are located in the main head and in a shelf off the main pillar. Class
II stars are scattered all over the region with no particular
association to the cloud. The Class I stars recently formed in the
pillar, and considering that they are much younger than the HII region,
they could have been triggered.

\cite{getman07} observed IC 1396N, another bright rim in the same
region, in x-ray and found an age sequence that suggests triggering
from south to north, into the rim. There are class III and class II
stars around the rim and class I/0 stars inside. \cite{beltran09} did a
JHK survey of the same bright rim and found few NIR-excess sources and
no signs of clustering toward the southern part of the rim. They also
found no color or age gradient in the north-south direction. They
concluded there was no triggering but perhaps there was a gradient in
the erosion of gas around protostars. \cite{choudhury10} observed the
region with Spitzer IRAC and MIPs and suggested there was an age
sequence with the younger stars in the center of the bright rim and the
older stars near the edge. They derived a propagation speed into the
rim of $0.1-0.3$ km s$^{-1}$.

The pillars of the Eagle Nebula, M16, are among the most famous cloud
structures suggestive of triggering. Several young stars appear at the
tips. It is difficult to tell if these stars were triggered by the
pressures that made the pillars, or if they existed in the head regions
before the HII pressure swept back the periphery. Triggering requires
that the pillar stars are much younger than the other stars in the
region.  Some exposure of existing stars could be possible if there is
a wide range of ages among the pillar and surrounding stars.

\cite{sugitani02} found Type I sources near the pillar heads in M16 and
older sources all around the pillars. They suggested there was an age
sequence within the pillar. \cite{fukuda02} observed M16 in $^{13}$CO,
C$^{18}$O, and 2.7 mm emission, finding a high density molecular core
at the end of the pillar, as expected from HII region compression.
However, \cite{indebetouw07} suggested that the young objects in the
area are randomly distributed and not triggered. They showed the
distribution of protostars in various stages of accretion and saw no
clear patterns with age. Thus the issue of triggering in the M16
pillars seems unresolved.

\cite{guar10} found a different age sequence in M16: the stars in the
northwest part of the whole HII region are younger than the stars in
the southeast part. They suggested that a 200 pc shell triggered both
M16 and M17 3 Myr ago on much larger scales.

IC 5146 is a filamentary cloud with low-level star formation at one end
\citep{lada99}. It is not in an HII region and the source of the
structure and pressure to shape it is not evident. It looks like it was
formerly a diffuse cloud that was compressed at one end by a supernova.
The ends of a filamentary cloud are the most susceptible parts to this
kind of random disturbance.

\cite{hoso07} studied molecule formation in compressed shells around
HII regions and suggested that H$_2$ could form without bright CO
emission during the expansion of the shell into a cold neutral medium.
They found an example of this in the shell around the W3-4-5 region.
There is cold HI and perhaps unobserved H$_2$, without any evident CO.
They proposed that this is an intermediate stage in the collapse of a
swept up shell and the site for future triggered star formation.

\begin{figure}[b]
% \vspace*{-2.0 cm}
\begin{center}
 \includegraphics[width=4.in]{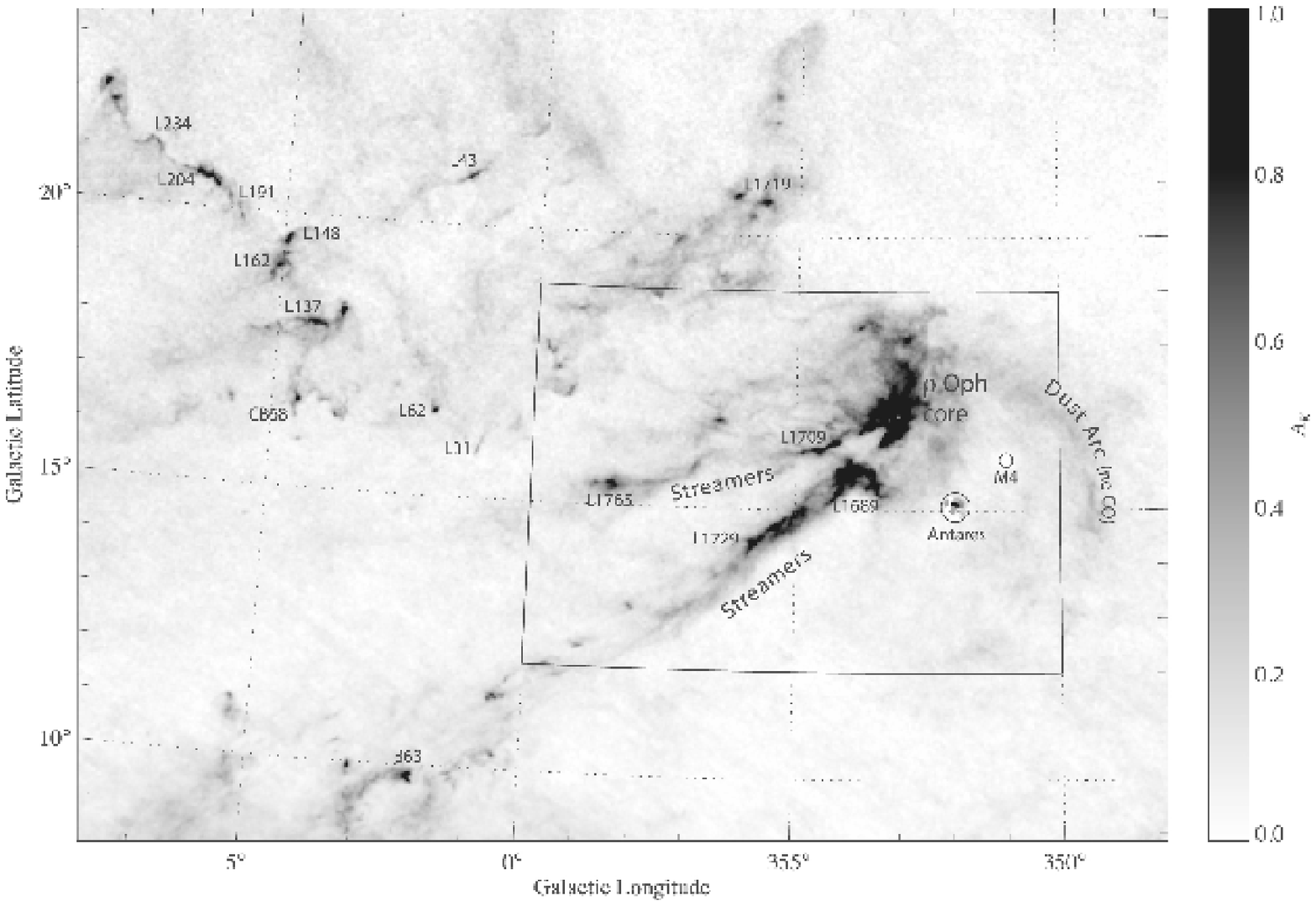}
% \vspace*{-1.0 cm}
\caption{Large-scale dust map of the Ophiuchus region, from
\cite{lombardi08}.}\label{lombardi}
\end{center}
\end{figure}

The Ophiuchus cloud core was swept back by pressures from the Sco-Cen
association \citep{degeus92}. Star formation in the rho Oph region
could have been triggered at the same time. A large-scale dust map of
the whole region is in Lombardi et al. (2008; see Figure
\ref{lombardi}). There are many protostars and dense cores
\citep[e.g.,][]{kirk05} in what looks like a giant pillar. Sco-Cen is
off the field to the upper right.

The Carina nebula has many bright rimmed clouds and pillars that were
recently studied by \cite{smith10}. They note that the young stars lag
the bright rims, as if they were left behind in an advancing ionization
front of cloud destruction.

\section{Age Sequences in Bright Rims}

Several of the references above look for or find age sequences of stars
along the axis of a bright rim. \cite{sugitani95} pioneered this. At
first, such a sequence seems obvious because the HII region is
expanding, so position correlates with time. However, the bright-rim
heads are usually more massive than the stars and the stars should be
gravitationally attracted to the heads. If the head acceleration by gas
pressure is less than their internal acceleration by gravity, then the
embedded stars will get pulled along with the heads as the heads move.
There would be little exposure of the stars in that case. To accelerate
the head faster than the internal gravitational acceleration means that
the pressure difference between the front and the back side has to
exceed the internal gravitational acceleration multiplied by the head
column density. Both of these quantities are essentially the internal
self-gravitational binding energy of the head if the head is
virialized, and therefore also the internal turbulent energy density.
In that case, the head has to be accelerated faster than the square of
its turbulent velocity dispersion divided by its radius. Such a larger
acceleration could occur when the initially low-density gas is first
compressed into a comet head by the HII region. Once the entire head is
compressed and star formation occurs in the dense gas, the inside of
the head will be close to pressure equilibrium with the outside
ionization front, and the relative acceleration will decrease.

A description of such pressure equilibrium and the resulting
acceleration is given by \cite{bert90}. In their equation 5.10, they
write the ratio of the self-gravitational acceleration inside the head
to the acceleration of the whole head by the rocket effect as
\begin{equation}
{{g_{\rm grav}}\over{g_{\rm rocket}}}\sim2\left({{M_{\rm
cl}}\over{M_{\rm Jeans}}}\right)^{2/3}\end{equation} for a non-magnetic
head, and \begin{equation}{{g_{\rm grav}}\over{g_{\rm rocket}}}\sim
\left({{M_{\rm cl}}\over{M_{\rm \Phi}}}\right)^2\end{equation} for a
magnetic head. Here, $M_{\rm cl}$ is the cloud head mass, $M_{\rm
Jeans}=1.18\sigma_{\rm cl}^4/(G^3P_{\rm cl})^{1/2}$ and $M_{\rm
\Phi}=0.12\Phi/G^{3/2}$ are the critical (or maximum) cloud masses for
stability without and with a magnetic field, respectively; $\Phi$ is
the total magnetic flux in the cloud. The pressure at the ionized
surface of the cloud enters the expression for $M_{\rm Jeans}$ and is
\begin{eqnarray}
{{P_{\rm cl}}\over k}=2.45\times10^7[S_{\rm 49}/(R_{\rm cl,pc}R_{\rm
pc}^2)]^{1/2} \;{\rm cm}^{-3}K \; {\rm if}\;(\psi>10)\;\\
=1.65\times10^9(S_{\rm 49}/R_{\rm pc}^2)\;{\rm cm}^{-3}K \;{\rm
if}\;(0.3<\psi<10)
\end{eqnarray}
for dimensionless parameter $\psi$,
\begin{equation}
\psi=\alpha F_{\rm II}R_{\rm cl}/\sigma_{\rm II}^2 =
5.15\times10^4{{S_{\rm 49}R_{\rm cl,pc}}\over{R_{\rm
pc}^2}}.\end{equation} In these expressions, the recombination rate to
all but the ground state is $\alpha$, the incident ionizing flux is
$F_{\rm II}$ in photons cm$^{-2}$ s$^{-1}$, the ionizing luminosity is
S$_{\rm 49}$ in photons s$^{-1}$, the cloud radius is $R_{\rm cl,pc}$
in pc, the distance to the ionizing source is $R_{\rm pc}$, in pc, the
velocity dispersion in the cloud is $\sigma_{\rm cl}$, and the velocity
dispersion in the HII region is $\sigma_{\rm II}$.

These equations suggest that if the cloud head requires gross
instability for a star to form, i.e., $M_{\rm cl}>M_{\rm Jeans}$ or
$M_{\rm cl}>M_{\rm \Phi}$ (or, $M_{\rm cl}>M_{\rm Jeans}+M_{\rm \Phi}$
in \cite{mckee89}, then the internal gravitational acceleration in the
head is always greater than the rocket acceleration. Thus the stars
that form in the head should follow the head along as it accelerates
away from the HII region.  Why are the stars ``left behind'' in this
case?

Another consideration is that the side of the dense core facing the HII
region could be continuously peeled away by the ionization. The speed
of this peeling is determined by the incident flux. After pressure
equilibrium, a D-type ionization front enters the compressed neutral
gas; ``D'' stands for density-bounded, i.e., the ionizing radiation is
stopped by gas absorption \citep{spitzer78}. The speed of such a front
into the dense gas is $\sigma_{\rm cl}^2/[2\sigma_{\rm II}]$. Because
$\sigma_{\rm II}>>\sigma_{\rm cl}$, this D-front speed is always much
less than $\sigma_{\rm cl}$. The orbit speed of a newly formed star
inside the head is of order $\sigma_{\rm cl}$, however, for a
self-gravitating head. Thus the speed at which the ionized side of the
pillar gets peeled away is always much less then the embedded stellar
speed. Thus, stars should not be exposed by ionization either.

Evidently, for both the rocket effect and ablation by ionization, stars
forming in an unstable head should should continuously fall back into
the head center faster than the surface of the head moves away from the
source of ionization. This makes the exposure of young stars and their
age gradients difficult to understand. It could explain, however, why
age gradients are seldom obvious -- the triggered stars scatter around
the head by gravitational forces.

One solution to this problem is that the head is stable on a large
scale with $\sigma_{\rm cl}$ equal to some turbulent speed that is
larger than the sound speed, or perhaps with magnetic support, and yet
inside of the head, there are local dense clumps that are unstable in
the sense that their masses are larger than the thermal Jeans mass
after the magnetic field has diffused out. In these cases, it might be
possible that $M_{\rm core}>M_{\rm Jeans,thermal}$ for localized star
formation while at the same time $g_{\rm grav}<g_{\rm rocket}$ for
exposure of the star after it forms.

Another model of triggered star formation is that there is a
pre-existing pillar-like shape with multiple clumps aligned to the HII
region. Then the compression front moves along the pillar from clump to
clump, triggering gravitational instabilities as it goes. The exposure
of the stars would follow the erosion of each clump, compressed one
after another.

Of course it is possible that the stars near the head were not
triggered. A key observation for triggering will be the velocities of
the young stars near the head in comparison to the head velocity. If
the stars are moving much slower than the head, then they could have
formed before the compression and rocket-like acceleration.

Simulations by \cite{dale07b} of triggered star formation in numerous
dense clumps of a pre-existing molecular cloud indicate the difficulty
in distinguishing between stars that formed previously and were exposed
by clump ionization or motion, and stars that were triggered by the
ionization pressure. In this study, the additional amount of star
formation that was from the triggering alone was only $\sim 30$\%.

\section{Shell Expansion: Collect and Collapse}

\begin{figure}[b]
% \vspace*{-2.0 cm}
\begin{center}
 \includegraphics[width=3.in]{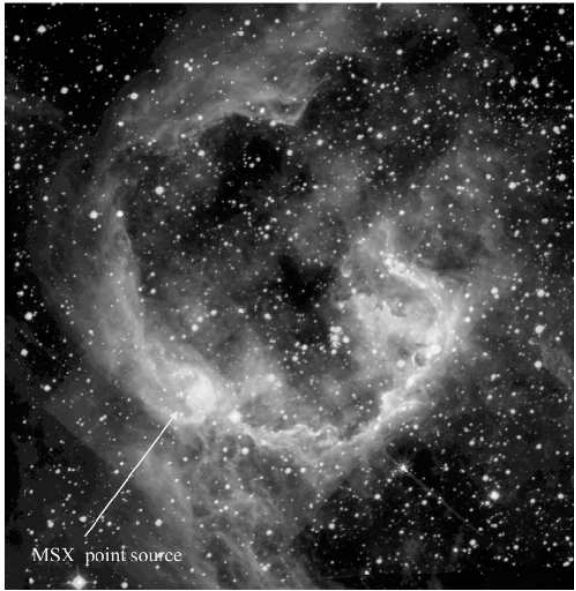}
% \vspace*{-1.0 cm}
\caption{Milky Way region W 79, consisting of a shell with dense clouds
and star formation at the edge, from Zavagno et al.
(2006).}\label{zavagno}
\end{center}
\end{figure}

\cite{zavagno06} studied star formation in the Milky Way source W 79
(Fig. \ref{zavagno}). It has a 1.7 Myr old shell with gravitationally
collapsed regions 0.1 Myr old along the perimeter. This is an example
of star formation triggering by the gravitational collapse of swept-up
gas around an older cluster or OB association. Sh2-219 is a similar
region \citep{deharveng06}. There is O9.5V star in a centralized HII
region, and a CO cloud, K-band embedded cluster, Ultra-compact HII
region, and Herbig Be star at the edge.

\cite{deharveng10} recently studied 102 bubbles and triggered star
formation using the Spitzer-GLIMPSE and MIPSGAL surveys for the IR, the
MAGPIS and VGPS surveys for the radio continuum, and the ATLASGAL
survey at 870 $\mu$m for cold dust emission. They found that 86\% of
the bubbles contain HII regions, and among those with adequate
resolution, 40\% have cold dust along their borders, presumably
accumulated during the bubbles' expansions. Eighteen bubbles have
either ultracompact HII regions or methanol masers in the peripheral
dust, indicating triggering.  They categorized their results into
several types of triggering. Star formation that occurs in pre-existing
cloud condensations is distinctive because the clouds protrude into the
bubble cavity like bright rims; 28\% of the resolvable shells are like
this. Star formation that occurs by the collapse of swept-up gas does
not protrude but is fully in the shell. That is because it is comoving
with the shell. In fact, clumps forming by gravitational collapse in a
shell could eventually protrude out of the front of the shell because
their higher column densities makes them decelerate slower than the
rest of the shell \citep{e89}. If this is observed, then the relative
position of a triggered clump and the shell around it should indicate
their relative speeds and the time when the clump first formed.

\cite{beerer10} studied Cygnus X North with Spitzer IRAC, classifying
stars according to their IR spectral ages. They found that younger
objects are in filaments that look compressed. Triggering of those
stars was suggested.

\cite{desai10} examined all 45 known supernova remnants in the LMC and
looked for associations with young stellar objects and with GMCs that
have no YSOs.   Seven SNR were associated with GMCs and YSOs, 3 SNRs
were with YSOs and no GMCs, and 8 SNRs were with GMCs and no YSOs. For
the 10 SNRs with YSOs, only 2 have YSOs that are clearly associated
with the SN shell, but in these cases, the SNe are younger than YSOs,
so the YSOs could not have been triggered. Desai et al. concluded that
SNe are too short-lived for triggering.

\section{General Aspects of Triggering}

Gas expands away from long-lived pressure sources like HII regions and
OB association bubbles. If the expansion scale is smaller than the
scale of a single cloud or the distance to a nearby cloud, then pillars
and bright rims form by the push-back of interclump gas. In this case,
star formation is a relatively fast process that works by squeezing the
pre-existing dense gas. The velocity of the triggered stars is smaller
than the overall shell expansion speed. The time delay between the
beginning of the pressure source and the formation of new stars is the
time for the pressure disturbance to reach the pre-existing cloud,
i.e., the HII region expansion time, plus the time for the pressure to
implode the cloud, which is relatively fast.

If the expansion scale is larger than the scale of a single cloud, then
shells form by the push-back of most nearby gas. A cavity then forms
with accumulated dense gas at the edge. This process of triggering is
relatively slow because new clumps have to form by gravitational
instabilities in the swept-up gas. The timescale for collapse and the
properties of shells when they collapse were investigated by
\cite{epe02} using collapse criteria in \cite{e94}. They ran several
thousand models of expanding shells in rotating, shearing galaxies and
found various trends with environmental factors. The basic time scale
for triggering was
\begin{equation} t\sim 4\xi^{1/10}(2\pi G\rho_{\rm
0})^{-1/2},\end{equation} where $\xi=\sigma^5/GL$ for source luminosity
$L$ and sound speed $\sigma$ in the shell, and for pre-shell density
$\rho_{\rm 0}$. They also found that the probability of collapse, or
the fraction of shells that collapse, $f$, depends on the Toomre $Q$
parameter for all possible variations in environment (see Lecture 1).
The relation is $f\sim0.5- 0.4\log_{10} Q$.

Simulations of shell formation and collapse around HII regions were
made by \cite{hoso05, hoso06a,hoso06b} and \cite{dale07a}.
\cite{hoso06a} found that shells driven into molecular clouds at
typical densities have time to fragment and form new stars. They showed
that at low ambient densities, the fragmentation can occur before CO
forms, but at high densities, the shell is primarily CO. \cite{hoso06b}
considered the minimum stellar mass that drives an expansion in which
triggered star formation produces a second generation mass comparable
to or larger than the first star; this stellar mass is around
$20\;M_\odot$ for a pre-shock density of 100 cm$^{-3}$. \cite{dale07a}
ran several simulations of an expanding HII region into a molecular
cloud and compared the resulting radii and times for collapse with the
analytical theory by \cite{whitworth94}, which represented the
numerical results well.

For the collect and collapse process, the velocity of triggered stars
in the swept-up region can be large, $\sim(P/\rho_{\rm 0})^{1/2}$ for
driving pressure $P$ and ambient density $\rho_{\rm 0}$.  Evidence for
triggering involves the causality condition: the triggering distance,
age difference, and relative velocity of the triggered stars compared
to the pressure-driving stars has to satisfy the relationship that the
distance equals the velocity times the time. The triggered stars have
to be much younger than the pressure-driving stars, and there has to be
a clear age bifurcation into triggering star ages and triggered star
ages, in order to be certain that triggering has occurred. Without a
clear age difference, the suspected triggered stars could be part of
the overall star formation process in the first generation, even if
they are located in a compressed clump near the edge of the pressurized
region.

Simulations of star formation triggered by ionization pressure have
been discussed by \cite{dale05, dale07b} and \cite{grit09}. These
simulations run for too short a time to generate an expanding coherent
shell and form stars by the collect and collapse mechanism. Triggering
instead is by the forced compression of pre-existing clumps. Because
stars are forming in these clumps anyway, the excess star formation
from triggering is small. Longer-time simulations could show more
triggering in the collect and collapse scenario. As mentioned above,
the timescale has to be several times the dynamical time in the
pre-shock material. This is a problem for clouds that are not
magnetically supported because they will collapse anyway in that time,
even without compression. Thus triggering, as observed in shells,
requires stability before the compression arrives, presumably from
magnetic forces, and instability after the compression, presumably from
enhanced magnetic diffusion in the compressed region combined with a
greater surface pressure for the given cloud mass.

Thick shells should differ from thin shells in their stability
properties because the gravitational forces in the shell are diluted by
thickness when it is large, as discussed for galactic disks in Section
1.1 of Lecture 1. \cite{wunsch10} studied the thick shell case for
shells that do not accumulate material as they expand, but are bound on
both sides by thermal pressure.

\section{ISM Energy Sources that may Trigger Star Formation}

There are many energy sources for the ISM but only a few are likely to
trigger star formation. The essential condition is that the energy
source has to change a cloud from stable to unstable. Usually this
requires some kind of compression, the compressed mass has to exceed an
unstable mass, and the compressive force has to last for a time
comparable to the collapse time in the compressed region. As mentioned
above, individual supernova seem too short-lived to trigger star
formation in the ambient ISM, even though they play an important role
in energizing the ISM.  More important are the HII regions, stellar
winds bubbles, and multiple supernovae that occur in OB associations
and star complexes.  Another long-range and long-lived source of
compression is a spiral density wave.

\section{Summary}

As we have seen over the last four lectures, star formation can be
initiated by a variety of processes, including spontaneous
gravitational instabilities in the combined stellar and gaseous medium,
occasional cloud collisions, especially in density-wave shocks, and
triggered gravitational instabilities in compressed regions ranging
from spiral-arm dust lanes, to Lindblad resonance rings, tidal arcs
around interacting galaxies, gaseous shells and rings in galactic
disks, and molecular clouds at the edges of HII regions. The role of
compression is either to bring some amount of otherwise stable gas
together so it can collapse and form stars on its own, or to compress
an existing cloud from a stable configuration to an unstable
configuration, at which point it, too, forms stars on its own. Always
accompanying this gas redistribution or compression is an enhancement
in internal energy dissipation. Otherwise, the region would have
collapsed into stars on its own.  Compression through a shock does this
reduction, or compression-enhancement of magnetic diffusion, or even
compression to reduce the turbulent dissipation time. Without
compression, the region may still collapse on its own, but with a
longer time scale.

Because the final step in all of these triggering scenarios is the
formation of stars deep in a cloud core, away from any pressure source
that may be acting on the cloud surface, the detailed processes of star
formation, such as stellar collapse, accretion, disk formation, and so
on, should not depend much on triggering. If the source of compression
also heats the gas, then perhaps the thermal Jeans mass increases in
the compressed region, and this might affect the stellar initial mass
function. However, higher driving luminosities and therefore higher
cloud temperatures are usually accompanied by higher pressures in a way
that the thermal Jeans mass stays about constant \citep{ekw08}.

The overall affect of triggered star formation on the average star
formation rate seems to be small in the main parts of galaxy disks. The
empirical laws discussed in Lecture 1 seem not to depend on how the
self-gravitating molecular gas is made, as long as it is made quickly
between previous molecular cloud disruptions. If the dispersed gas from
a previous event of star formation lingers around in a diffuse state
for a long time, without forming stars, then it might still turn
molecular from self-shielding, thereby contributing to $\Sigma_{\rm
H2}$, but not contribute in the right proportion to $\Sigma_{\rm SFR}$.
The empirical \cite{bigiel08} law would then fail. We suggested in
Lecture 2 that this may be the case for dust clouds in the interarm
regions of M51, i.e., that they are marginally stable to have lasted so
long from their formation in the previous spiral arm. But the fraction
of molecules in a non-gravitating form cannot be large for the
correlation between $\Sigma_{\rm SFR}$ and $\Sigma_{\rm H2}$ to work
out as well as it does. Frequent gas compression by all of the various
pressures in the ISM, combined with the forced loss of internal energy
that accompanies this compression, ensures that most of the molecular
and atomic debris from one event of star formation soon makes it into
another event of star formation. Prevalent triggering thereby acts as a
scavenger for inert diffuse clouds, keeping most parts of the ISM in a
constant state of collapse or imminent collapse. This is the saturation
in star formation that previous lectures have mentioned.

With very low star formation rates, as in dwarf galaxies and the outer
parts of disks, a much higher fraction of the gas can be in diffuse
form, and then triggering can play a more direct role in the average
star formation rate. At a very minimum, it can provide locally high
pressures where the thermal stability of the gas allows a cool phase to
exist in equilibrium with the radiation field. Without such cool
phases, disk instabilities will just make warm and diffuse flocculent
spirals in the gas, and there will not be enough dense matter to affect
the star formation rate. Put simply, at very low average pressures,
cool diffuse clouds seem to require pressure disturbances for their
formation from the warm phase. Most commonly, outer spiral arms seem to
do this, but stellar pressure sources might be important too. This
enablement of cool cloud formation is presumably the first step in the
condensation process that leads to star formation.

%%-----------------------------
%%      your bibliography
%%-----------------------------

\end{document}